\documentclass[journal]{IEEEtran}

\usepackage{graphicx}
\usepackage{amsmath,amssymb}
\usepackage{makecell}
\usepackage{booktabs}
\usepackage{multirow}
\usepackage{array}
\usepackage{cite}
\usepackage{url}
\usepackage{xcolor}
\usepackage{subcaption}
\usepackage{microtype}
\usepackage{newtxtext,newtxmath}
\usepackage[hidelinks]{hyperref}
\AtBeginDocument{%
  }

\usepackage{float}
\usepackage{microtype}
\usepackage{graphicx}
\usepackage{booktabs}
\usepackage{makecell}
\begin{document}

\title{Joint-Embedding Predictive Architecture for Solar PV Panel Fault Classification}

\author{
Seyyedhamid Azimidokht,
Mehdi Monemi,
Abdelhak Kharbouch,
Farid Hamzehaghdam,
Mehdi Rasti, Jamshid Aghaei and Emil Kurvinen%
\thanks{Seyyedhamid Azimidokht, Mehdi Monemi, Abdelhak Kharbouch, Farid Hamzehaghdam, Mehdi Rasti, and Emil Kurvinen are with the University of Oulu, Oulu, Finland. E-mails: \{seyyedhamid.azimidokht, mehdi.monemi, abdelhak.kharbouch, farid.hamzehaghdam, mehdi.rasti, emil.kurvinen\}@oulu.fi. Jamshid Aghaei is with Central Queensland University, Australia. E-mail: j.aghaei@cqu.edu.au.}
}

\maketitle
\captionsetup{labelfont=bf,textfont=normalfont}
\begin{abstract}
The rapid expansion of solar photovoltaic (PV) systems has increased the need for reliable and scalable fault classification, as manual inspection is impractical at scale. Thermal infrared (IR) imaging provides a non-contact solution for identifying PV faults; however, accurate classification remains challenging due to class imbalance, limited texture information, and subtle thermal differences. In this work, we investigate the applicability of Joint-Embedding Predictive Architecture (JEPA) for thermal IR PV fault classification across various scenarios and propose JEFFNet (JEPA-EFFicientNet), a multibranch architecture that combines JEPA-based self-supervised representation learning with EfficientNetV2-S-based supervised convolutional feature extraction. JEFFNet fuses semantic representations from a JEPA-pretrained Vision Transformer with convolutional features from EfficientNetV2-S, enabling complementary feature learning. JEFFNet is evaluated on two public thermal IR datasets, PVF-10 and InfraredSolarModules (ISM), for both multiclass and derived binary (healthy/faulty) classification. On PVF-10, JEFFNet achieves an F1-score of $93.21$ and an accuracy of $94.33$ in the 10-class task, and an F1-score of $97.53$ and an accuracy of $96.41$ in the derived 2-class task. On ISM, JEFFNet achieves an F1-score of $72.60$ and an accuracy of $83.88$ in the 12-class task, and an F1-score of $94.69$ and an accuracy of $94.78$ in the derived 2-class task. JEFFNet also uses only 108.6M parameters versus 205.91M for GEPFNet, a 47.2\% reduction. These results demonstrate that combining self-supervised semantic and supervised convolutional features provides an effective, parameter-efficient solution for thermal IR PV fault classification. \footnote{The source code is publicly available at \url{https://github.com/Azimi2kht/JEFFNet}} 
\end{abstract}

\begin{IEEEkeywords}
Joint-Embedding Predictive Architecture (JEPA), photovoltaic (PV) panel, fault classification, infrared  thermography, EfficientNetV2, self-supervised learning, solar PV inspection.
\end{IEEEkeywords}

\section{Introduction}

Global renewable electricity capacity is projected to grow rapidly between 2025 and 2030, with solar photovoltaic (PV) systems expected to drive most of this expansion \cite{IEA2025Renewables}. As PV deployment increases at large scale, traditional manual inspection and conventional monitoring become insufficient for ensuring reliability, performance, and safety. Recent reports by the IEA and IEA-PVPS highlight that AI- and ML-based methods can improve fault detection, predictive maintenance, and operational efficiency, helping reduce outages and maintenance costs \cite{IEA2025EnergyAI,IEAPVPS2025Trends}. In parallel, IEC standards such as IEC TS 62446-3:2017 provide important imaging protocols for PV inspection and support the development of automated computer vision pipelines \cite{IEC62446-3-2017,JRC127142}.

PV farms are exposed to harsh environmental conditions, including high irradiation, temperature variations, humidity, and mechanical stress, all of which contribute to degradation and diverse fault types over time \cite{WAQARAKRAM2022118822}. Early fault detection is therefore essential, since undetected failures can reduce power generation, accelerate degradation, and create safety risks at the module and string levels \cite{IEAPVPS2017Failures}. Consequently, automated PV fault detection and classification has become an important task for large-scale operation and maintenance, especially in utility-scale installations where manual inspection is impractical.

Several sensing modalities have been explored for PV inspection, including visible-light (RGB) imaging, infrared (IR) thermography, electroluminescence (EL) imaging, ultraviolet (UV) imaging, and electrical measurements. Among these, IR thermographic imaging has emerged as a key technology for PV inspection. IR imaging enables the non-contact visualization of thermal anomalies associated with electrical and structural faults, such as hotspots, short circuits, and connection failures, even under varying illumination conditions. Its importance is formally recognized in the IEC TS 62446-3:2017 standard \cite{IEC62446-3-2017}, which specifies the use of IR thermography as a recommended technique for condition monitoring and fault diagnosis of PV modules and strings. By capturing temperature distributions rather than visible textures, IR images provide fault-relevant information that is often imperceptible in RGB imagery, making them particularly suitable for early-stage fault detection. In this work, we focus on IR-based PV fault classification. Most existing works on IR-based PV fault analysis rely on fully supervised learning or handcrafted feature engineering, and therefore do not explicitly exploit self-supervised representation learning or semantic prediction in latent space. As a result, the learned features are often tied to low-level appearance patterns and may be less robust when discriminative cues are subtle, as is common in thermal IR imagery. In this context, recent advances in representation learning have shown that Joint Embedding Predictive Architecture (JEPA) provides an effective alternative for high-level visual understanding tasks. 
Unlike reconstruction-based methods that emphasize pixel fidelity or contrastive methods that depend on pairwise alignment, JEPA learns by predicting target embeddings from context embeddings in a shared latent space. This encourages the model to capture semantically meaningful, task-relevant structure while suppressing irrelevant low-level variations, making it appealing for PV fault classification from IR images, where textures are limited and defect evidence is often subtle. 

Recent studies indicate that JEPA is not limited to generic representation-learning benchmarks, but can also be effective for downstream classification tasks across different modalities. In particular, Assran et al. \cite{assran2023self} demonstrated its effectiveness in image classification. However, existing JEPA-based classification studies have employed a single-branch setting, where only one feature-extraction pathway is used. A multibranch or fusion architecture refers to a framework in which multiple feature-extraction pathways process the same input in parallel and their learned representations are combined before final classification. In this work, we propose a novel multibranch framework that fuses JEPA-based self-supervised representations with EfficientNetV2-S features for thermal IR PV fault classification. The motivation for this fusion is that the two branches capture complementary aspects of the same input image: JEPA provides semantic latent representations learned through self-supervised prediction, while EfficientNetV2-S provides discriminative convolutional features optimized for classification. By combining these representations, the proposed framework is expected to produce a richer image representation than either branch alone.

\subsection{Related work}
PV fault analysis has been studied mainly using three imaging modalities: RGB, electroluminescence (EL), and Infrared (IR). In RGB imagery, recent work has used deep learning for RGB defect detection and classification; for example, Authors of \cite{zhu5022044novel} proposed a recent RGB-based defect detector, while Authors of \cite{ramachandran2025deep} showed that preprocessing and class balancing can materially affect RGB-based fault classification performance. However, RGB inspection is mainly limited to surface-visible defects and is sensitive to image quality, illumination, and dataset realism. In EL imaging, defect analysis has also been extensively studied because EL images reveal fine cell-level structural defects. Authors of \cite{deitsch2019automatic} addressed automatic classification of defective PV cells in EL images, while Authors of \cite{fioresi2021automated} proposed semantic-segmentation-based defect detection and localization in EL images and introduced the UCF EL Defect dataset. Nevertheless, EL inspection typically requires controlled acquisition conditions and electrical excitation, which makes it less practical for large-scale field deployment. Aligned with many of existing related studies, our work is based on IR imagery, which is particularly attractive for PV fault analysis because it directly captures temperature distributions associated with operational faults and is therefore well suited for practical field-scale inspection.

Several studies have investigated PV fault analysis using IR thermography due to its ability to directly capture thermal anomalies associated with operational defects. Authors of \cite{dotenco2016automatic} proposed an automatic pipeline for aerial IR inspection of PV plants that first detects individual PV modules and then identifies thermal abnormalities, such as overheated modules, hot spots, and overheated substrings, using statistical analysis of IR imagery. Authors of \cite{ali2020machine} proposed a machine-learning framework for hotspot identification in PV modules using IR thermography, where handcrafted hybrid features, including RGB, texture, histogram of oriented gradient (HOG) and local binary pattern (LBP) descriptors, were fused and classified with a support vector machine (SVM). Their method categorized PV panels into healthy, non-faulty hotspot, and faulty classes, achieving 96.8\% training accuracy and 92\% testing accuracy while maintaining relatively low computational complexity and storage requirements. 

Authors of \cite{dunderdale2020photovoltaic} investigated PV defect classification from thermal IR images using both feature-based machine learning and deep learning methods. In particular, they evaluated SIFT-based descriptors with conventional classifiers alongside VGG-16 and MobileNet, showing that deep CNN models achieved the strongest multiclass defect-classification performance, with MobileNet reaching 89.5\% accuracy. Authors of \cite{le2023thermal} proposed a lightweight CNN-based framework for thermal inspection of PV modules and showed that accurate anomaly classification can be achieved in real time on edge devices. Their approach reached 85.35\% accuracy on a 12-class IR dataset. Authors of \cite{le2021remote} employed a ResNet-based convolutional neural network combined with ensemble learning for PV anomaly detection and classification. Evaluated on a dataset of 20{,}000 IR images spanning 12 defect categories, their method achieved 94.4\% accuracy for anomaly detection and 85.9\% for multi-class fault classification. Authors of \cite{kellil2023fault} investigated PV fault diagnosis from IR images using both a lightweight DCNN and a fine-tuned VGG-16 model. Their results showed that transfer learning can achieve very high performance for both binary fault detection and multiclass fault diagnosis in thermographic PV inspection. Recent studies have also explored Vision Transformers (ViT)\cite{dosovitskiy2020image} for thermal PV fault detection. Transformer-based models such as ViT-Tiny and Swin-Tiny have shown strong binary and multiclass classification performance \cite{aksoy2025benchmarking}; however, some fault categories, particularly environmentally induced defects such as soiling, remain difficult to distinguish, highlighting limitations related to thermal image quality and subtle inter-class differences. More recently, authors of  \cite{GUO2026104014} proposed GEPFNet, a group-equivariant feature extraction and parallel fusion network for fault classification. By combining group-equivariant feature learning with parallel feature fusion, GEPFNet achieved state-of-the-art results, with an F1-score of 92.47\% and accuracy of 94.64\% on the 10-class task, and an F1-score of 95.01\% and accuracy of 96.05\% in the healthy-versus-faulty 2-class setting.

Meanwhile, JEPA has recently shown strong potential for downstream tasks such as classification, by learning compact semantic representations self-supervisedly \cite{monemi2025tutorial}. By predicting target embeddings from context embeddings in latent space rather than reconstructing raw inputs, JEPA provides transferable features that have been shown to be effective for supervised classification. For example, I-JEPA demonstrated strong performance in downstream image classification task using representations learned in latent space, while Stochastic Positional Embeddings-JEPA (StoP-JEPA) further improved I-JEPA-style masked image modeling by introducing stochastic positional embeddings to reduce overfitting to exact target locations and improve robustness to location uncertainty \cite{bar2023stochastic}. A-JEPA \cite{fei2023jepa} extended the JEPA paradigm to audio and speech classification and reported state-of-the-art results across multiple benchmarks. Similarly, Authors of \cite{weimann2025self} showed that JEPA-based pretraining improves downstream classification performance on electrocardiogram (ECG) data.

\subsection{Motivations and Contributions}

Although recent studies have shown that deep learning models can achieve strong performance for IR-based PV fault classification, the literature still has two important gaps. First, most existing methods rely on fully supervised training or handcrafted features, and therefore do not exploit self-supervised semantic representation learning in latent space. This is a notable limitation for IR PV imagery, where textures are weak and the visual differences between fault categories can be subtle. Second, while CNN-based supervised models are effective at learning discriminative task-specific patterns, the combination of such supervised representations with JEPA-based self-supervised semantic representations within the same modality has not been explored. Motivated by these limitations, the main contributions of this work are listed as follows:

\begin{itemize}
    \item JEPA learns semantically meaningful latent representations by predicting target embeddings from context embeddings, making it well suited to IR PV images where discriminative cues are subtle and low-level textures are limited. In this work, we use the context encoder obtained from StoP-JEPA training for PV panel fault classification. Specifically, we first jointly train the context encoder, target encoder, and predictor in the the StoP-JEPA framework on the PVF-10 dataset, a widely used PV fault dataset in a self-supervised manner, and then use the trained context encoder as a feature extractor by attaching a classifier head and performing linear probing. Since JEPA is built upon a Vision Transformer (ViT) encoder, whose representation quality generally benefits from large-scale and semantically rich pretraining data, we also investigate the effect of transferring semantic knowledge learned from a broader visual domain; We evaluate a context encoder pretrained in the StoP-JEPA setting on ImageNet-1k \cite{5206848}, a large-scale semantically rich dataset, and adapt it to the PV fault classification task using the same linear probe strategy. The use of semantically rich out-of-domain ImageNet-1k pretraining improves the JEPA representations, leading to a significant improvement in downstream PV fault classification performance.

    \item Building on the best-performing linear-probing setting, we further improve the adaptation of the StoP-JEPA context encoder using a staged fine-tuning strategy. In this strategy, the pretrained context encoder is first kept frozen and only the classifier head is trained, allowing the randomly initialized head to adapt to the PV fault classification task without disturbing the learned representations. The encoder is then unfrozen and the whole network is fine-tuned end-to-end. On the PVF-10 dataset, the {\it staged} fine-tuning strategy achieves 93.86\% accuracy and 93.69\% F1-score. This substantially improves over two extreme {\it non-staged} training strategies: (a) linear-probing setting, which achieves 85.02\% accuracy and 82.70\% F1-score, as well as the setting in which the JEPA-encoder and the classifier head are trained together from the beginning, which achieves 88.09\% accuracy and 82.41\% F1-score.

   \item To further enhance the performance, we introduce JEFFNet, an architecture that combines JEPA embeddings with EfficientNetV2-S features. The motivation for this design is to exploit the complementary strengths of the two components: JEPA provides semantic and transferable latent representations, while EfficientNetV2-S provides strong supervised local and task-specific discriminative features. JEFFNet adopts the staged training strategy described above, where the pretrained StoP-JEPA context encoder is first kept frozen while EfficientNetV2-S and the classifier head are trained, then the full network is fine-tuned end-to-end. We show that this framework improves over the staged single-branch StoP-JEPA classifier and achieves strong performance on two widely used PV fault  datasets, PVF-10 and ISM. On the PVF-10 10-class setting, JEFFNet achieves an F1-score of 93.21\% and recall of 93.19\%, improving over GEPFNet \cite{GUO2026104014}, the recent state-of-the-art network for PV fault classification, by 0.74\% in F1-score and 0.78\% in recall. On the ISM 12-class benchmark, JEFFNet achieves an F1-score of 72.60\% and recall of 71.47\%, improving over GEPFNet by 1.28\% in F1-score and 2.66\% in recall.
    
   \item In addition to the original multiclass settings, we also study 2-class scenarios in which all fault categories are grouped into a single \textit{defective} class and contrasted against the \textit{healthy} class. This setting is particularly important in practice because it directly reflects the screening task of deciding whether a PV panel is faulty or not. On PVF-10, JEFFNet achieves an F1-score of 97.53\% and recall of 97.49\%, improving over GEPFNet by 2.52\% and 2.48\%, respectively. On ISM, JEFFNet achieves an F1-score of 94.69\% and recall of 93.16\%, improving over GEPFNet by 0.76\% in F1-score, while also achieving notable gains in precision and MCC of 2.34\% and 1.69\%, respectively.
   
\end{itemize}

\section{Methodology}
\label{sec:method}
We first provide some prelimiaries in section~\ref{preliminaries}. Then we introduce the framework designed for classification of the solar panel defects.

\subsection{Preliminaries}
\label{preliminaries}
This section briefly introduces the two main building blocks underlying the proposed framework: the JEPA paradigm to enable  self-supervised semantic representation learning, and EfficientNetV2-S, which provides efficient convolutional feature extraction. These preliminaries establish the foundation for the proposed architectures presented in the next section.

\subsubsection{Joint Embedding Predictive Architecture (JEPA) Image Implementations: I-JEPA and StoP-JEPA}

\begin{figure*}[!t]
    \centering
    \includegraphics[width=\linewidth]{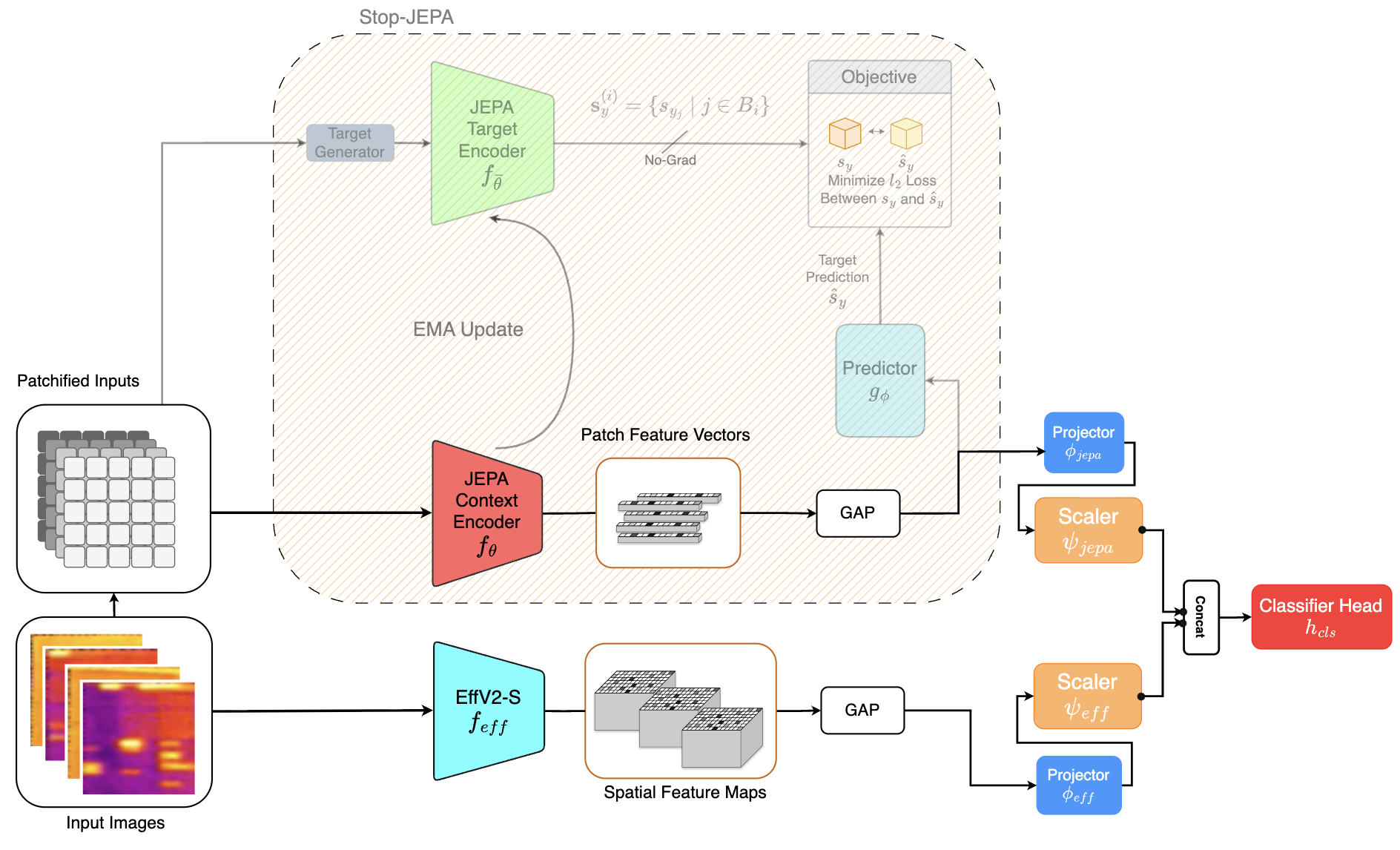}
    \caption{Architecture of the proposed JEFFNet fusion model. The input thermal IR image is processed by two parallel branches: a JEPA-pretrained ViT encoder, which produces patch-level semantic feature vectors that are aggregated by global average pooling, and an EfficientNetV2-S backbone, which extracts spatial convolutional feature maps followed by global average pooling. The resulting branch representations are transformed through projection and scaling modules, concatenated, and passed to the classifier head to predict the PV fault class.}
    \label{fig:jeffnet}
\end{figure*}

JEPA is a self-supervised learning framework that learns representations by predicting the latent embeddings of unseen targets from a visible context, rather than reconstructing the input signal itself. This latent prediction objective encourages the model to capture semantic and task-relevant structure while avoiding excessive emphasis on low-level image details. I-JEPA \cite{assran2023self} is the first image-based implementation of the JEPA paradigm. The orange hatched region in Fig.~\ref{fig:jeffnet} illustrates the I-JEPA-style pretraining mechanism. Given an input image, the image is first divided into a sequence of non-overlapping patches and a masking strategy is applied on the patch grid. Let $B_i$ denote the set of patch indices corresponding to the $i$-th target block, for $i=1,\dots,M$, and let $B_x$ denote the visible context indices. The masked views are processed by two encoder branches: a context encoder $f_{\theta}$ and a target encoder $f_{\bar{\theta}}$, both typically implemented as Vision Transformers (ViTs). The context encoder processes the visible context patches, whereas the target encoder provides the target patch representations used as prediction targets. The target encoder is updated through an exponential moving average (EMA) of the context encoder parameters. This slowly evolving target encoder provides stable target representations and helps prevent trivial solutions. The target encoder produces patch-level representations
\begin{equation}
\mathbf{s}_y = f_{\bar{\theta}}(y)
= \{s_{y_1}, s_{y_2}, \dots, s_{y_N}\},
\end{equation}
where $s_{y_j}$ is the embedding of the $j$-th patch. For each target block $B_i$, the corresponding target representation is
\begin{equation}
\mathbf{s}_y^{(i)} = \{s_{y_j} \mid j \in B_i\}.
\end{equation}
Similarly, the context encoder produces the visible context representation
\begin{equation}
\mathbf{s}_x = f_{\theta}(x)
= \{s_{x_j} \mid j \in B_x\}.
\end{equation}
A predictor network $g_{\phi}$ then uses the encoded context together with positional embedding about the target locations to produce the target predictions $\hat{\mathbf{s}}_y^{(i)}$. The model is trained by minimizing the discrepancy between the predicted and target representations over all target blocks:
\begin{equation}
\mathcal{L}
=
\frac{1}{M}
\sum_{i=1}^{M}
D\bigl(\hat{\mathbf{s}}_y^{(i)}, \mathbf{s}_y^{(i)}\bigr)
=
\frac{1}{M}
\sum_{i=1}^{M}
\sum_{j \in B_i}
\left\|
\hat{s}_{y_j} - s_{y_j}
\right\|_2^2 .
\end{equation}
For downstream image-level classification, only the pretrained context encoder is retained as the JEPA feature extractor, while the target encoder and predictor are discarded. StoP-JEPA extends I-JEPA by addressing location uncertainty in masked image modeling. Instead of conditioning the predictor on deterministic positional embeddings of the target patches, StoP-JEPA perturbs the target positional embeddings with Gaussian noise before they are provided to the predictor. Specifically, for the positional embedding $p_j$ associated with the $j$-th target patch, StoP-JEPA uses the positional embedding
\begin{equation}
\tilde{p}_j = p_j + \epsilon_j,\qquad
\epsilon_j \sim \mathcal{N}(\mathbf{0}, \sigma^2\mathbf{I}),
\end{equation}
where $\sigma$ controls the noise magnitude. The predictor $g_{\phi}$ then receives the encoded context together with the positional embeddings $\{\tilde{p}_j\}$ to generate the target predictions. By introducing uncertainty into the target positions during training, StoP-JEPA discourages the predictor from relying on exact spatial locations and instead encourages the encoder to learn more semantically meaningful representations. Consequently, StoP-JEPA preserves the latent-space prediction objective of I-JEPA while improving the transferability of the learned representations.

\subsubsection{EfficientNetV2}
EfficientNetV2 is a family of convolutional neural networks designed to improve training speed and parameter efficiency while maintaining strong classification performance \cite{tan2021efficientnetv2}. It builds upon EfficientNet \cite{tan2019efficientnet}, which introduced compound scaling of network depth, width, and input resolution. EfficientNetV2 further improves this design by using \textit{Fused-MBConv} blocks in the early stages, where the expansion and depthwise convolution are combined into a regular convolution, while retaining MBConv blocks with squeeze-and-excitation (SE) modules in deeper stages \cite{hu2018squeeze}. This design provides an efficient balance between representational capacity, model size, and training speed.

\begin{figure}[!t]
    \centering
    \includegraphics[width=\linewidth]{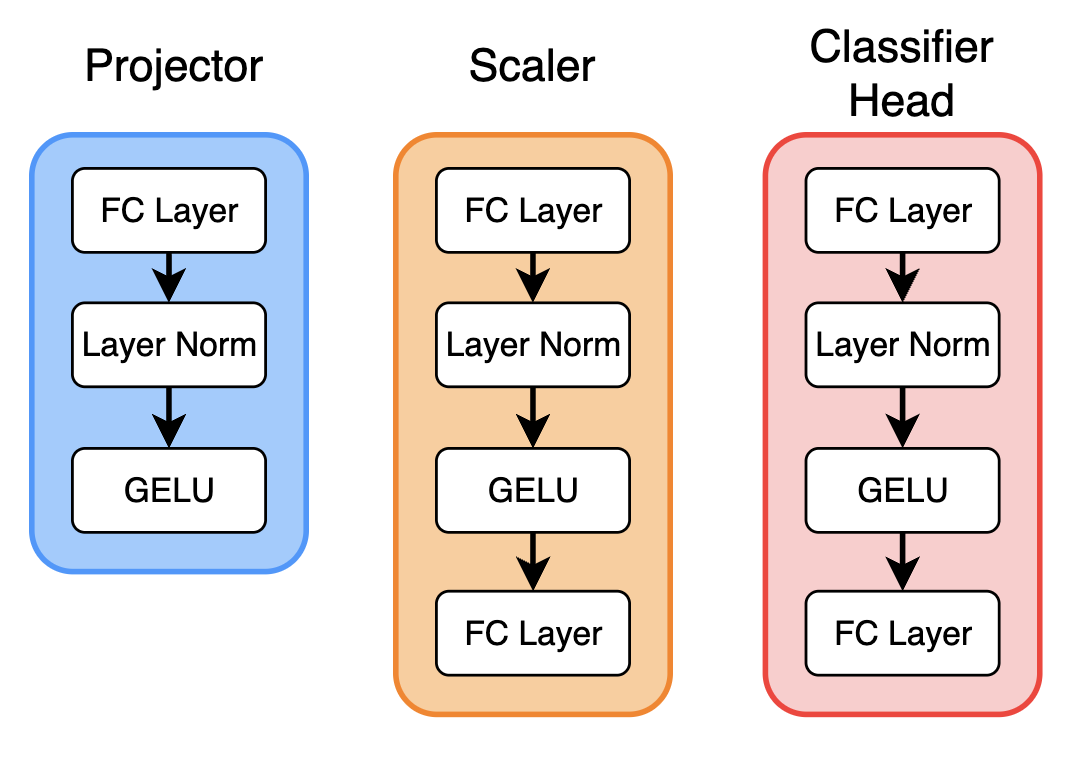}
    \caption{Detailed structure of the projection, scaling, and classifier modules used in JEFFNet. The projector maps each branch representation into a shared feature space using a fully connected layer followed by layer normalization and GELU activation. The scaler further refines the projected representation using two fully connected layers with intermediate normalization and activation. The classifier head maps the fused representation to the final class logits.}
    \label{fig:blocks}
\end{figure}

\subsection{JEFFNet}

As illustrated in Fig.~\ref{fig:jeffnet} and Fig.~\ref{fig:blocks}, JEFFNet is designed for thermal IR PV fault classification by combining StoP-JEPA representations with EfficientNetV2-S features. In the StoP-JEPA pretraining, the context encoder, target encoder, and predictor has been trained in a self-supervised manner on ImageNet-1k \cite{5206848}, a large-scale semantically rich dataset. For the downstream classification task, only the pretrained context encoder is retained, while the target encoder and predictor are deactivated because no masked-target prediction is performed during classification. To provide complementary discriminative information, an EfficientNetV2-S branch is added alongside the StoP-JEPA context encoder, followed by a classifier head. During classification, the whole input image is fed to the StoP-JEPA context encoder and, in parallel, to EfficientNetV2-S. The extracted representations are projected into a shared feature space, refined through branch-specific transformation modules, concatenated, and passed to the classifier head for fault prediction. JEFFNet is trained using a staged strategy in which the pretrained StoP-JEPA context encoder is first kept frozen while the classifier head and EfficientNetV2-S branch are trained, and then the context encoder is unfrozen so that the full network can be fine-tuned end-to-end.

Let $x \in \mathbb{R}^{H \times W \times C}$ denote an input IR image. In the JEPA branch, the input is processed by the pretrained JEPA context encoder,
\begin{equation}
\mathbf{s}_{x} = f_{\theta}(x),
\end{equation}
where $\mathbf{s}_{x}=\{s_{x_j}\}_{j=1}^{N}$ denotes the sequence of patch-level latent representations, $N$ is the number of image patches, and $s_{x_j} \in \mathbb{R}^{d_{\mathrm{jepa}}}$ is the embedding of the $j$-th patch produced by the JEPA context encoder. Since the target task is image-level classification, the patch-level representations are aggregated using global average pooling:
\begin{equation}
z_{\mathrm{jepa}} =
\frac{1}{N}
\sum_{j=1}^{N} s_{x_j}.
\end{equation}
The resulting vector $z_{\mathrm{jepa}}$ is then mapped to a shared projection space using a projection module,
\begin{equation}
p_{\mathrm{jepa}} = \phi_{\mathrm{jepa}}(z_{\mathrm{jepa}}),
\end{equation}
where $\phi_{\mathrm{jepa}}(\cdot)$ consists of a fully connected linear layer, layer normalization, GELU activation, and dropout. The projected representation is further processed by a branch-specific transformation module,
\begin{equation}
\tilde{z}_{\mathrm{jepa}} = \psi_{\mathrm{jepa}}(p_{\mathrm{jepa}}),
\end{equation}
where $\psi_{\mathrm{jepa}}(\cdot)$ consists of a fully connected linear layer, layer normalization, GELU activation, dropout, and a final fully connected linear layer. In parallel, the same input image is passed through an EfficientNetV2-S backbone. The classification layer of the backbone is removed, and global average pooling is used to obtain a convolutional feature vector:
\begin{equation}
z_{\mathrm{eff}} = f_{\mathrm{eff}}(x),
\end{equation}
where $z_{\mathrm{eff}} \in \mathbb{R}^{d_{\mathrm{eff}}}$ denotes the feature representation extracted by EfficientNetV2-S. This feature vector is projected into the same shared representation space:
\begin{equation}
p_{\mathrm{eff}} = \phi_{\mathrm{eff}}(z_{\mathrm{eff}}),
\end{equation}
where $\phi_{\mathrm{eff}}(\cdot)$ follows the same structure as the JEPA projection module. The projected convolutional representation is then refined using a transformation module:
\begin{equation}
\tilde{z}_{\mathrm{eff}} = \psi_{\mathrm{eff}}(p_{\mathrm{eff}}),
\end{equation}
where $\psi_{\mathrm{eff}}(\cdot)$ has the same structure as the JEPA transformation module.

The two transformed representations are concatenated:
\begin{equation}
z_{\mathrm{fused}} =
\left[
\tilde{z}_{\mathrm{eff}}
\; \Vert \;
\tilde{z}_{\mathrm{jepa}}
\right],
\end{equation}
where $\Vert$ denotes feature concatenation. The resulting representation is then passed to the classifier head:
\begin{equation}
\hat{y} = h_{\mathrm{cls}}(z_{\mathrm{fused}}),
\end{equation}
where $\hat{y} \in \mathbb{R}^{C}$ denotes the output logits and $C$ is the number of target classes. The classifier head consists of a fully connected linear layer, layer normalization, GELU activation, dropout, and a final fully connected linear layer. The model is optimized using weighted cross-entropy loss to account for class imbalance. For a batch of size $B$, the loss is defined as
\begin{equation}
\mathcal{L}
=
-
\frac{1}{B}
\sum_{i=1}^{B}
w_{y_i}
\log
\frac{
\exp(\hat{y}_{i,y_i})
}{
\sum_{c=1}^{C} \exp(\hat{y}_{i,c})
},
\end{equation}
where $y_i$ is the ground-truth label of the $i$-th sample, $\hat{y}_{i,c}$ is the predicted logit for class $c$, and $w_{y_i}$ is the class weight assigned to the ground-truth class. The class weights are computed from the training subset using the standard balanced weighting strategy, where each class is weighted inversely proportional to its frequency in the training data.

\begin{figure*}[!t]
    \centering

    \begin{subfigure}[b]{0.47\textwidth}
        \centering
        \includegraphics[width=\linewidth]{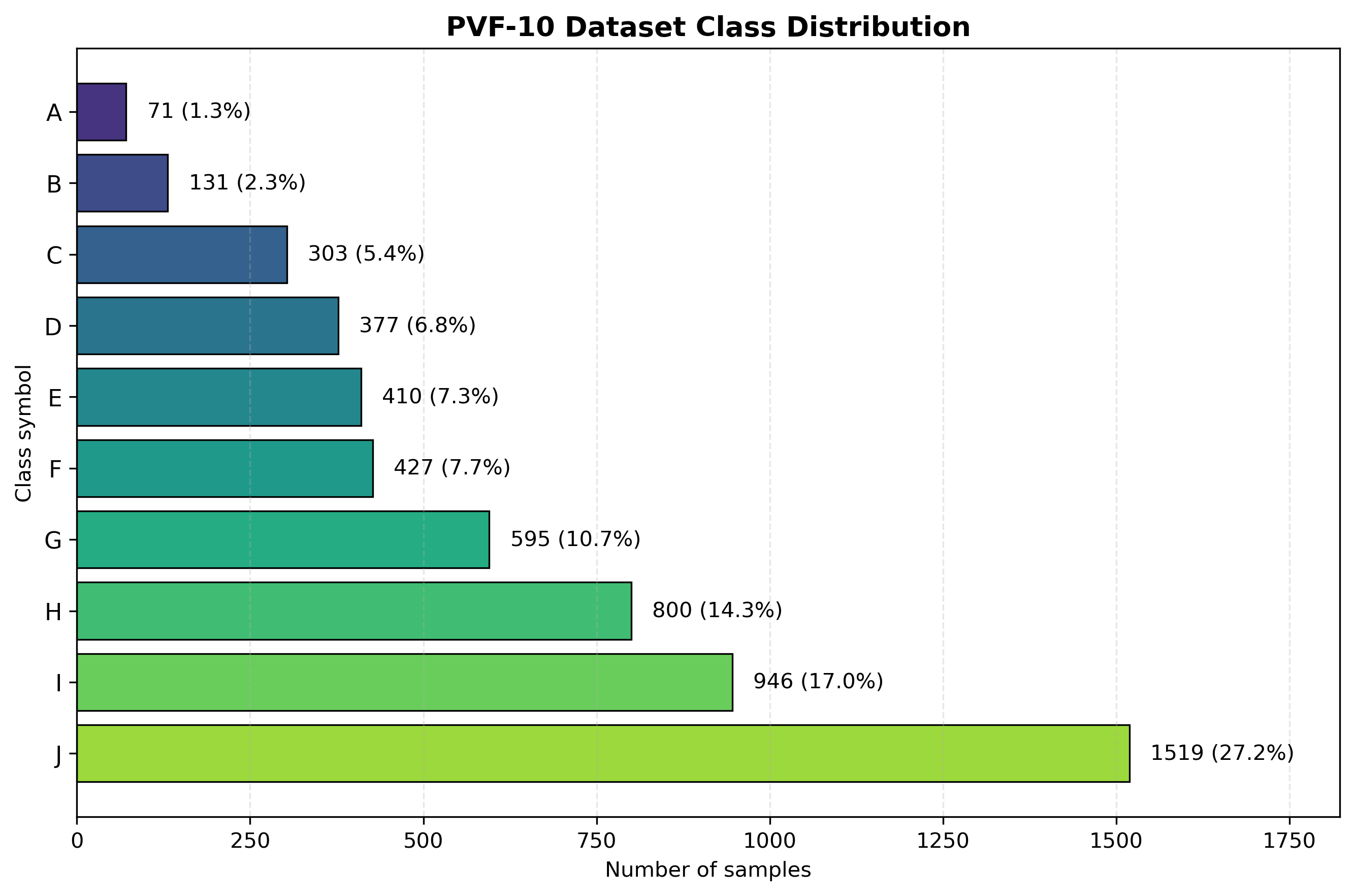}
        \caption{PVF-10 dataset.}
        \label{fig:pvf10_distribution}
    \end{subfigure}
    \hfill
    \begin{subfigure}[b]{0.47\textwidth}
        \centering
        \includegraphics[width=\linewidth]{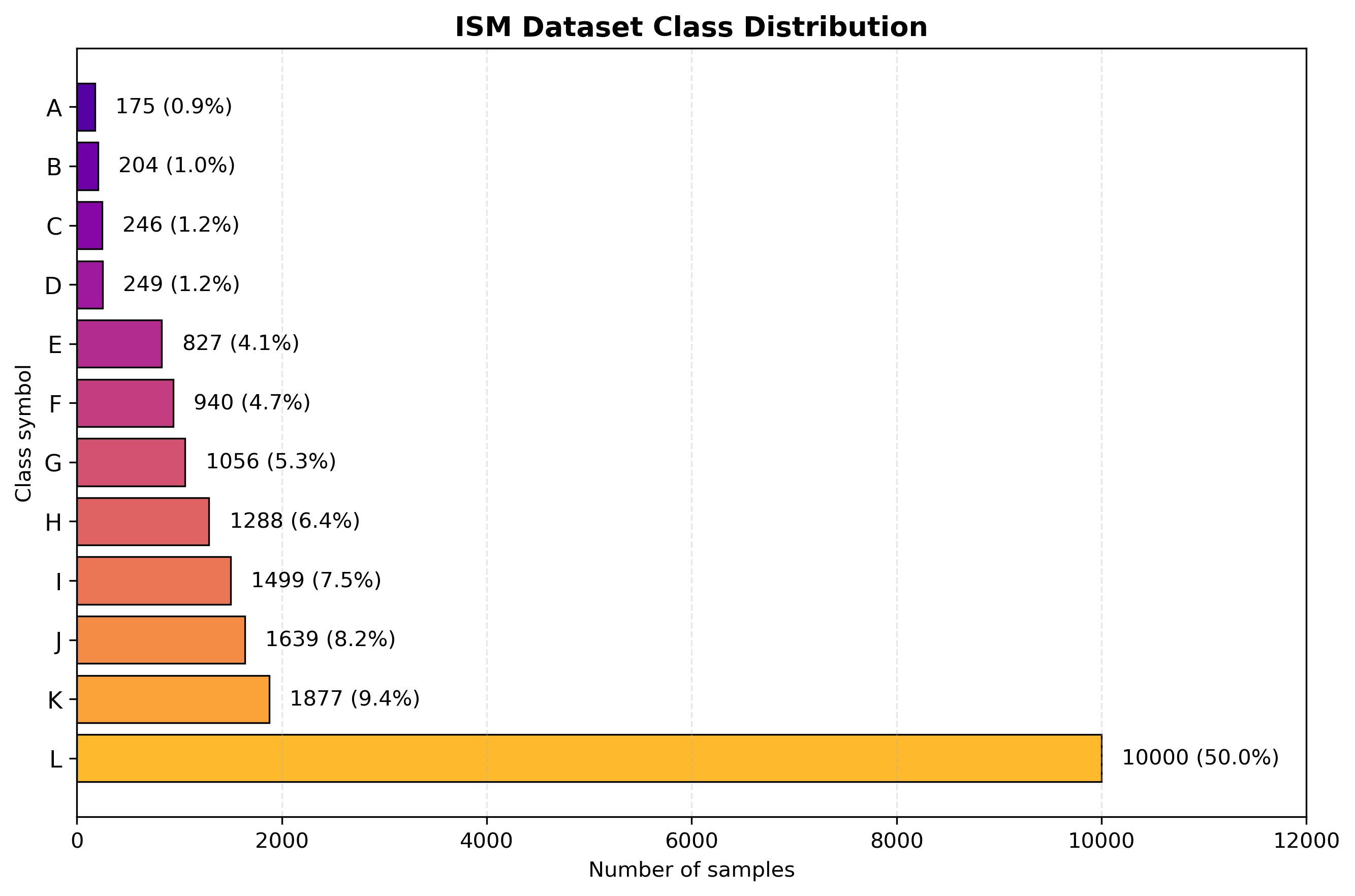}
        \caption{ISM dataset.}
        \label{fig:ism_distribution}
    \end{subfigure}

    \caption{Class distributions of the datasets. Subfigure (a) shows the PVF-10 dataset, where the class symbols are: A: string short circuit, B: break, C: shadow, D: bottom dirt, E: debris cover, F: short circuit panel, G: substring open circuit, H: junction box heat, I: hot cell, and J: healthy panel. Subfigure (b) shows the ISM dataset, where the class symbols are: A: Diode-Multi, B: Soiling, C: Hot-Spot-Multi, D: Hot-Spot, E: Offline-Module, F: Cracking, G: Shadowing, H: Cell-Multi, I: Diode, J: Vegetation, K: Cell, and L: No-Anomaly.}
    \label{fig:dataset_distribution}
\end{figure*}

\begin{figure}
    \centering
    \includegraphics[width=\linewidth]{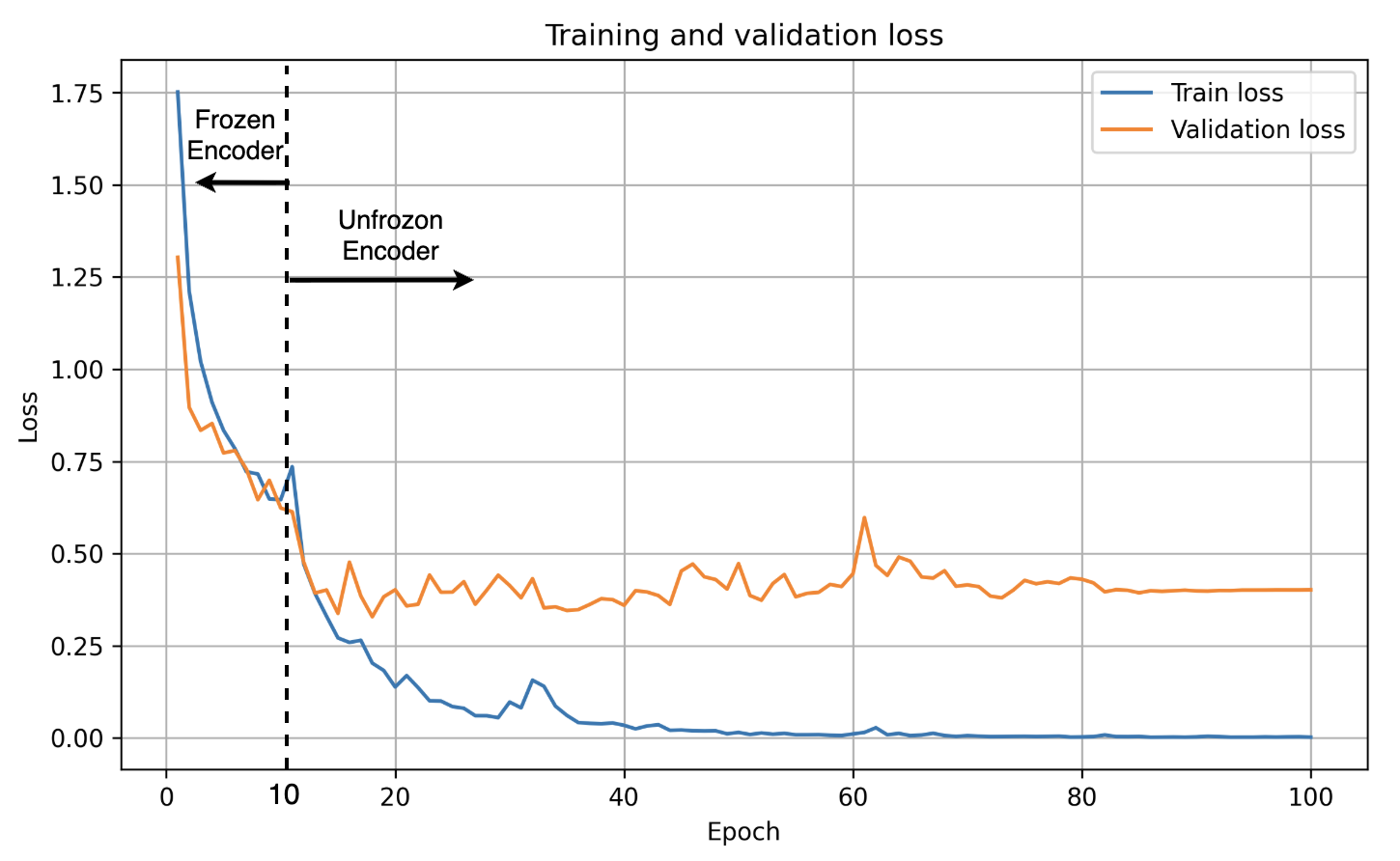}
    \caption{The training and validation loss curves for proposed model.}
    \label{fig:loss_curve}
\end{figure}

\section{Experiments}
\subsection{Experimental Setup}
\subsubsection{Datasets}
The proposed method was evaluated on two publicly available datasets: PVF-10 \cite{wang2024pvf} and InfraredSolarModules (ISM) \cite{millendorf2020infrared}. PVF-10 is a IR dataset designed for fine-grained PV fault classification. It contains 5579 annotated TIR images collected by unmanned aerial vehicles (UAVs) from operational PV farms under diverse acquisition conditions. The dataset includes images from different PV plants and regions, with variations in latitude, weather, solar exposure, and time of day, resulting in differences in illumination, thermal radiation, and scene complexity. These properties make PVF-10 a realistic benchmark for evaluating fine-grained PV fault classification methods under low thermal contrast and complex outdoor conditions. The dataset is annotated into ten classes, including \textit{substring open circuit}, \textit{panel short circuit}, \textit{string short circuit}, \textit{debris covering}, \textit{shadowed panel}, \textit{bottom dirt}, \textit{broken panel}, \textit{junction box overheating}, \textit{overheated cell}, and \textit{healthy panel}. In this study, the 112$\times$112 version of PVF-10 was used for model evaluation. The ISM dataset contains 20,000 IR images of size $24\times40$ pixels collected from aerial inspections of operational solar farms. It comprises 12 classes, including 11 anomaly types \textit{Cell}, \textit{Cell-Multi}, \textit{Cracking}, \textit{Hot-Spot}, \textit{Hot-Spot-Multi}, \textit{Shadowing}, \textit{Diode}, \textit{Diode-Multi}, \textit{Vegetation}, \textit{Soiling}, and \textit{Offline-Module} together with a \textit{No-Anomaly} class. Fig.~\ref{fig:dataset_distribution} illustrates the class distributions of both datasets. As shown in the figure, both datasets exhibit noticeable class imbalance, with certain defect categories containing substantially more samples than others. This imbalance reflects the natural occurrence frequencies of PV faults in real-world installations and presents an additional challenge for learning robust fault classification models. For PVF-10, we followed the fixed training and test split provided with the dataset. From the original training split, 10\% of the samples were further separated as a validation set using a stratified random split to preserve the class distribution. The validation set was used for model selection, and the best-performing model on the validation set was then evaluated on the fixed test set. For the ISM dataset, since no predefined training, validation, and test split is provided, we constructed the splits from the full dataset. First, 10\% of the samples were held out as the test set using stratified sampling. From the remaining 90\%, 10\% was further separated as the validation set, again using stratified sampling to preserve the class distribution. The remaining samples were used for training. The validation set was used for model selection, and the selected best-performing model was evaluated once on the held-out test set. 

\subsubsection{Evaluation Metrics}
Since both PVF-10 and ISM datasets are imbalanced, we report macro-averaged
metrics to give equal importance to all fault classes. Let $C$ be the number
of classes, and let $\mathrm{TP}_i$, $\mathrm{FP}_i$, $\mathrm{TN}_i$, and
$\mathrm{FN}_i$ denote the true positives, false positives, true negatives,
and false negatives for class $i$, respectively. Overall accuracy is
computed over all samples,
\begin{equation}
    \mathrm{Acc} =
    \frac{\sum_{i=1}^{C}\mathrm{TP}_i}{N},
\end{equation}
where $N$ is the total number of samples. Precision, recall, and F1-score
are first computed per class and then averaged,
\begin{align}
    \mathrm{Pre}_i &= \frac{\mathrm{TP}_i}{\mathrm{TP}_i+\mathrm{FP}_i}, \quad
    \mathrm{Re}_i = \frac{\mathrm{TP}_i}{\mathrm{TP}_i+\mathrm{FN}_i}, \quad
    \mathrm{F1}_i = \frac{2\,\mathrm{Pre}_i\,\mathrm{Re}_i}{\mathrm{Pre}_i+\mathrm{Re}_i}, \\
    \mathrm{Pre} &= \frac{1}{C}\sum_{i=1}^{C}\mathrm{Pre}_i, \quad
    \mathrm{Re} = \frac{1}{C}\sum_{i=1}^{C}\mathrm{Re}_i, \quad
    \mathrm{F1} = \frac{1}{C}\sum_{i=1}^{C}\mathrm{F1}_i.
\end{align}
For AUC, we use the macro-averaged one-vs-rest formulation,
\begin{equation}
\mathrm{AUC} =
\frac{1}{C}\sum_{i=1}^{C}\mathrm{AUC}_i,
\end{equation}
where $\mathrm{AUC}_i$ is the area under the ROC curve when class $i$
is treated as positive and all remaining classes as negative. We also
report the multiclass Matthews correlation coefficient (MCC), computed
directly from the confusion matrix $\mathbf{M}\in\mathbb{R}^{C\times C}$
following Gorodkin~\cite{gorodkin2004comparing},
\begin{equation}
\mathrm{MCC} =
\frac{
c \cdot s - \sum_{k=1}^{C} p_k t_k
}{
\sqrt{
\left(s^2 - \sum_{k=1}^{C} p_k^2\right)
\left(s^2 - \sum_{k=1}^{C} t_k^2\right)
}
},
\end{equation}
where $s=\sum_{i,j}M_{ij}$ is the total number of samples,
$c=\sum_{k}M_{kk}$ is the number of correctly classified samples,
$t_k=\sum_{i}M_{ki}$ is the number of times class $k$ truly occurs, and
$p_k=\sum_{i}M_{ik}$ is the number of times class $k$ is predicted.

\begin{figure}[!h]
  \centering
  \includegraphics[width=\linewidth]{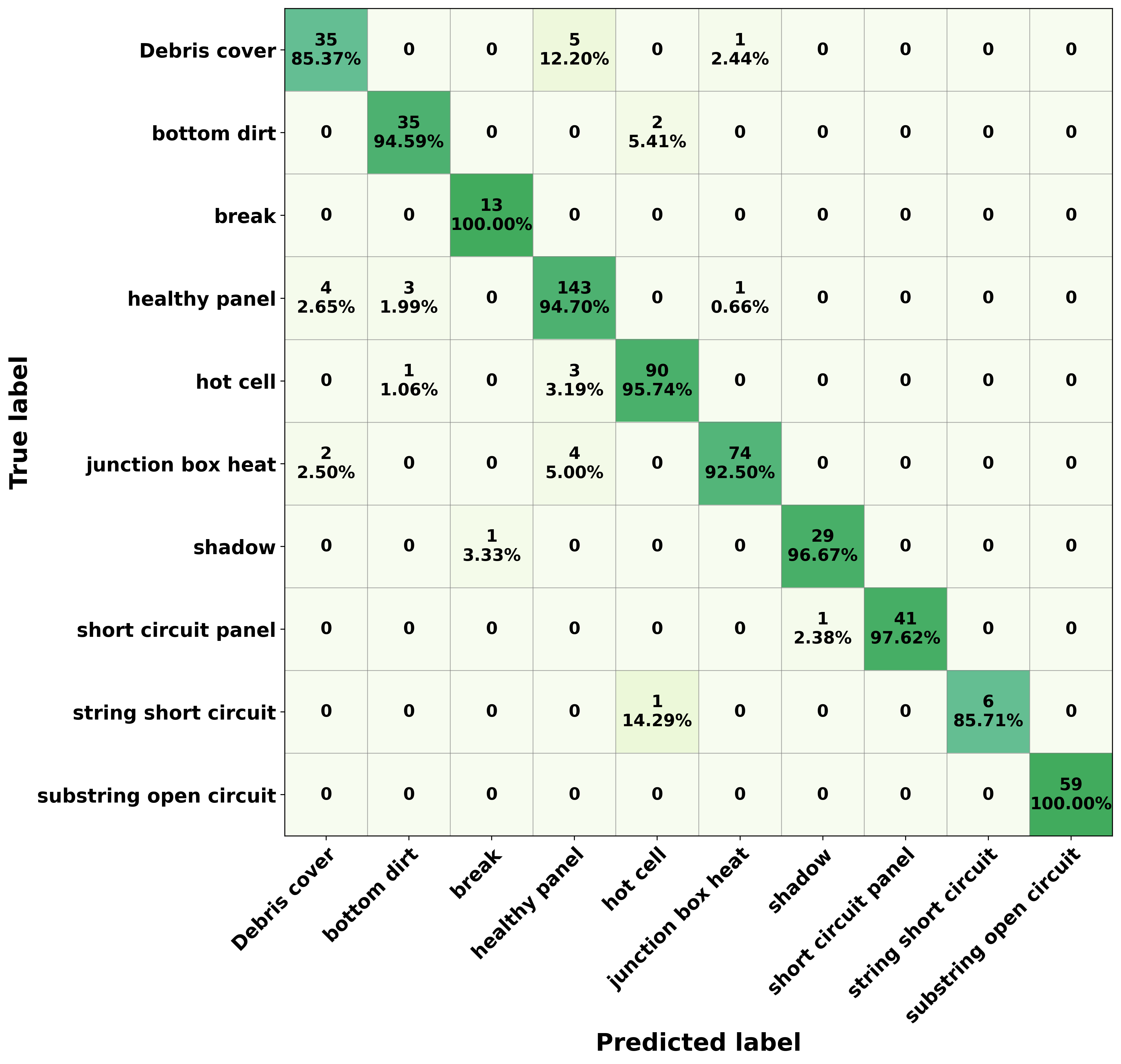}
  \caption{Confusion matrix of the best-performing JEFFNet run on the PVF-10 test set for the 10-class classification task. Each cell shows the sample count and corresponding row-wise percentage, with color intensity indicating the class-wise prediction rate.}
  \label{fig:cm-best}
\end{figure}

\begin{figure}[!h]
  \centering
  \includegraphics[width=\linewidth]{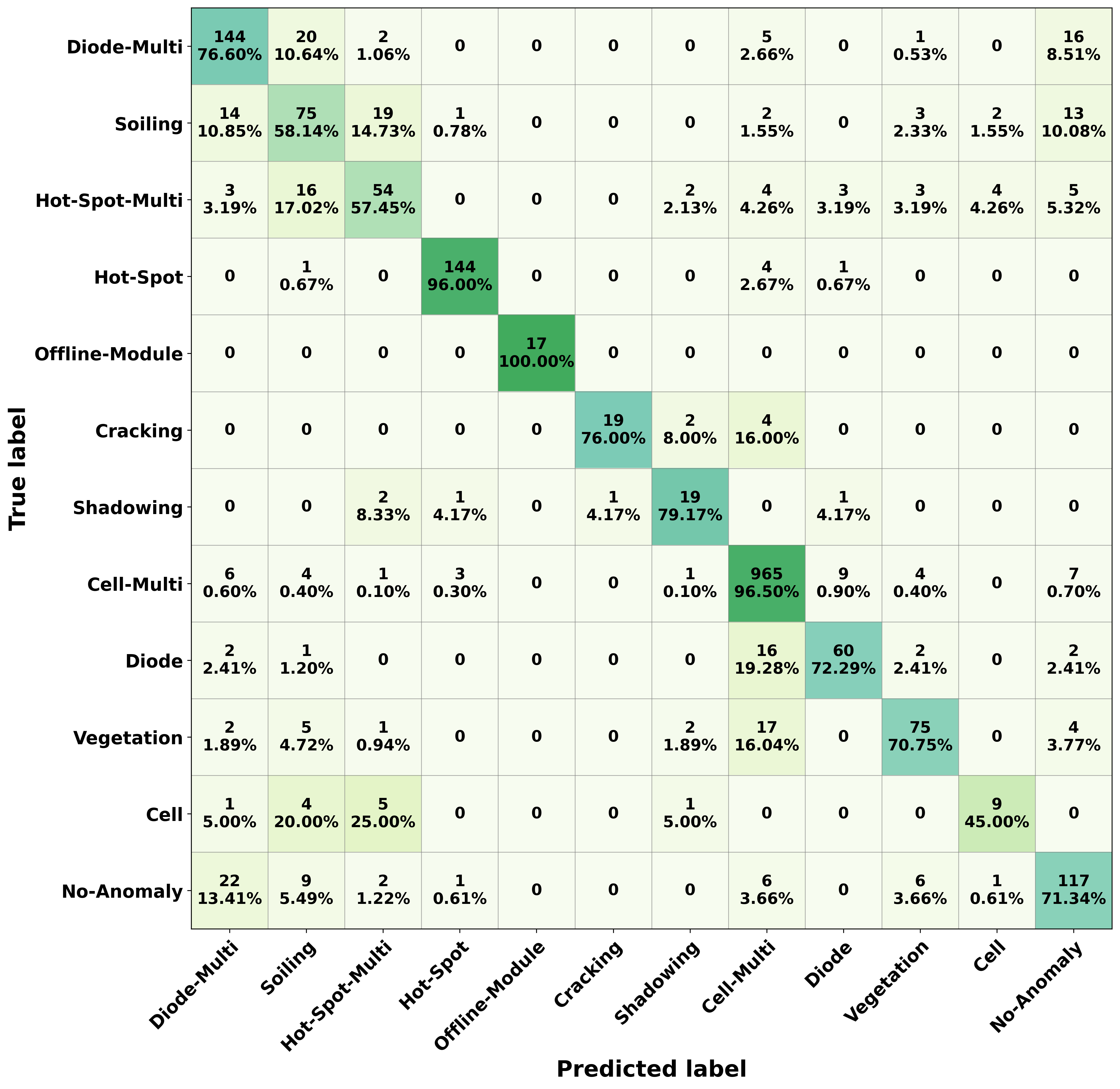}
  \caption{Confusion matrix of the best-performing JEFFNet run on the ISM dataset for the 12-class classification task. Each cell shows the sample count and corresponding row-wise percentage, with color intensity indicating the class-wise prediction rate.}
  \label{fig:ism-cm-best}
\end{figure}

\begin{table*}[!t]
\centering
\caption{Ablation study of the proposed JEFFNet components on the PVF-10 10-class task. The results in this table are obtained from a single representative run under the same fixed train/validation/test split and are intended only to analyze the contribution of different architectural and fine-tuning components.}
\label{tab:ablation}
\begin{tabular}{clccc ccccccc}
\toprule
No. & Encoder & \makecell{Pretraining JEPA \\ Dataset} & Training Strategy & Multibranch & Acc & F1 & Re & Pre & AUC & MCC \\
\midrule
1 & ViT-B (StoP-JEPA) & PVF-10 & Frozen & $\times$ & 68.95 & 65.56 & 67.21 & 65.97 & 93.16 & 63.51 \\

2 & ViT-B (StoP-JEPA) & ImageNet-1k & Frozen & $\times$ & 85.02 & 82.70 & 87.10 & 80.46 & 98.53 & 82.56 \\

3 & ViT-B (StoP-JEPA) & ImageNet-1k & Unfrozen & $\times$ & 88.09 & 82.41 & 82.16 & 82.92 & 98.41 & 85.91 \\

4 & ViT-B (StoP-JEPA) & ImageNet-1k & Staged & $\times$ & 93.86 & 93.69 & 92.98 & 94.58 & 99.69 & 92.73 \\
\midrule
5 & ViT-B (StoP-JEPA) + EfficientNetV2-S & ImageNet-1k & Staged & $\checkmark$ & $\mathbf{94.77}$ & $\mathbf{93.84}$ & $\mathbf{92.34}$ & $\mathbf{95.86}$ & $\mathbf{99.70}$ & $\mathbf{93.80}$ \\
\bottomrule
\end{tabular}
\end{table*}

\subsection{Implementation Details}

All input images were resized to $224 \times 224$ pixels before being fed to the model. For the JEPA branch, we used a ViT-B/16 encoder, where each image is divided into non-overlapping patches of size $16 \times 16$. The JEPA encoder was initialized from the ImageNet-1k pretrained StoP-JEPA ViT-B checkpoint released by the authors of StoP-JEPA\cite{bar2023stochastic}.
 For downstream PV fault classification, we adopted a staged fine-tuning strategy. During the training, data augmentation was applied to improve robustness and reduce overfitting. The augmentations included random horizontal flipping with probability $0.5$, random vertical flipping with probability $0.5$, and brightness jittering with a brightness factor of $0.5$. Images were normalized using the ImageNet mean and standard deviation. For validation and testing, only resizing and normalization were applied. The proposed JEFFNet was trained using a staged fine-tuning strategy for a total of $100$ epochs. During the first $10$ epochs, the pretrained ViT encoder was kept frozen while the EfficientNetV2-S branch, the feature fusion modules, and the classification head were trained. The ViT encoder was then unfrozen, and the entire network was jointly optimized for the remaining $90$ epochs. The learning rate was set to $3 \times 10^{-4}$ for the newly initialized layers and $1 \times 10^{-5}$ for the pretrained JEPA encoder after it was unfrozen. The training loss curves are shown in Fig.~\ref{fig:loss_curve}. All experiments were conducted using a batch size of $128$ on an NVIDIA A100-40GB GPU.

\begin{table*}[!t]
\centering
\caption{Comparison of existing methods and the proposed model on the PVF-10 dataset for the 10-Class task.}
\label{tab:pvf10-10-class}
\resizebox{\textwidth}{!}{%
\begin{tabular}{c l c c c c c c}
\toprule
\textbf{No.} & \textbf{Models/Study} & \textbf{F1} & \textbf{Acc} & \textbf{Re} & \textbf{Pre} & \textbf{AUC} & \textbf{MCC} \\
\midrule
1  & AlexNet \cite{GUO2026104014}
& $85.65 \pm 1.31$ & $90.31 \pm 1.26$ & $85.65 \pm 1.34$ & $85.94 \pm 1.49$ & $98.94 \pm 0.15$ & $88.53 \pm 1.48$ \\

2  & EfficientNetB0 \cite{GUO2026104014}
& $86.89 \pm 1.13$ & $91.53 \pm 0.90$ & $86.41 \pm 1.54$ & $87.69 \pm 0.83$ & $98.87 \pm 0.17$ & $89.97 \pm 1.07$ \\

3  & MobileNetV4 Small \cite{GUO2026104014}
& $86.43 \pm 2.46$ & $91.12 \pm 0.83$ & $85.90 \pm 2.81$ & $87.32 \pm 2.27$ & $98.70 \pm 0.30$ & $89.48 \pm 0.99$ \\

4  & ConvNeXtV2 Base \cite{GUO2026104014}
& $83.62 \pm 2.10$ & $89.31 \pm 1.10$ & $82.96 \pm 2.59$ & $84.88 \pm 1.70$ & $98.66 \pm 0.31$ & $87.34 \pm 1.31$ \\

5  & FastViT SA \cite{GUO2026104014}
& $88.08 \pm 1.53$ & $92.24 \pm 0.63$ & $88.17 \pm 1.43$ & $88.42 \pm 1.63$ & $98.96 \pm 0.27$ & $90.82 \pm 0.73$ \\

6  & ResNet50 \cite{wang2024pvf}
& $89.78$ & $92.42$ & $89.02$ & $89.34$ & -- & -- \\

7  & EfficientNetV2 Small \cite{wang2024pvf}
& $91.33$ & $93.14$ & $89.86$ & $90.38$ & -- & -- \\

8  & Coat Lite Small \cite{wang2024pvf}
& $90.95$ & $93.32$ & $87.46$ & $88.74$ & -- & -- \\

9  & Swin Transformer V2 Tiny \cite{wang2024pvf}
& $85.85$ & $89.71$ & $84.53$ & $85.05$ & -- & -- \\

10 & Vision Transformer Small \cite{wang2024pvf}
& $78.36$ & $83.57$ & $76.29$ & $77.06$ & -- & -- \\

11 & GEPFNet \cite{GUO2026104014}
& $92.47 \pm 0.87$ 
& \boldmath$94.64 \pm 0.35$
& $92.41 \pm 1.46$ 
& $92.72 \pm 0.71$ 
& $99.39 \pm 0.14$ 
& \boldmath$93.66 \pm 0.42$ \\

\midrule
12 & \textbf{JEFFNet{} (Ours)}
& \boldmath$93.21 \pm 1.24$
& $94.33 \pm 0.79$
& \boldmath$93.19 \pm 1.18$
& \boldmath$93.66 \pm 2.08$
& \boldmath$99.65 \pm 0.10$
& $93.30 \pm 0.92$ \\
\bottomrule
\end{tabular}%
}
\vspace{2mm}

{\footnotesize\raggedright
\textit{Note:} Bold values indicate the best results. In the 10-Class task on the PVF-10 dataset, our method and the comparative methods provide the average and standard deviation over 10 runs, while the results of the compared methods reported by \cite{wang2024pvf} and \cite{GUO2026104014} are taken directly from the paper. ``--'' indicates that the metric is not available.\par}
\end{table*}

\begin{table*}[!t]
\centering
\caption{Comparison of existing methods and the proposed model on the PVF-10 dataset for the 2-Class healthy vs faulty detection task.}
\label{tab:pvf10-2-class}
\resizebox{\textwidth}{!}{%
\begin{tabular}{c l c c c c c c}
\toprule
\textbf{No.} & \textbf{Models/Study} & \textbf{F1} & \textbf{Acc} & \textbf{Re} & \textbf{Pre} & \textbf{AUC} & \textbf{MCC} \\
\midrule
1  & AlexNet \cite{GUO2026104014} &
$94.03 \pm 0.82$ & $95.25 \pm 0.64$ & $94.21 \pm 1.14$ & $93.91 \pm 0.88$ & $98.36 \pm 0.18$ & $88.11 \pm 1.62$ \\

2  & EfficientNetB0 \cite{GUO2026104014} &
$93.65 \pm 0.83$ & $94.95 \pm 0.65$ & $93.81 \pm 0.95$ & $93.51 \pm 0.86$ & $98.46 \pm 0.38$ & $87.32 \pm 1.66$ \\

3  & MobileNetV4 Small \cite{GUO2026104014} &
$93.50 \pm 0.73$ & $94.75 \pm 0.62$ & $94.32 \pm 0.65$ & $92.79 \pm 0.94$ & $98.55 \pm 0.13$ & $87.09 \pm 1.42$ \\

4  & ConvNeXtV2 Base \cite{GUO2026104014} &
$93.53 \pm 0.96$ & $94.84 \pm 0.76$ & $93.78 \pm 1.18$ & $93.31 \pm 1.02$ & $98.31 \pm 0.27$ & $87.09 \pm 1.94$ \\

5  & FastViT SA \cite{GUO2026104014} &
$92.00 \pm 1.07$ & $93.55 \pm 0.91$ & $92.67 \pm 0.87$ & $91.42 \pm 1.36$ & $97.89 \pm 0.38$ & $84.07 \pm 2.10$ \\

6  & ResNet50 \cite{wang2024pvf} &
$93.68 \pm 0.47$ & $95.04 \pm 0.36$ & $93.30 \pm 0.72$ & $94.12 \pm 0.58$ & $98.53 \pm 0.30$ & $87.40 \pm 0.94$ \\

7  & EfficientNetV2 Small \cite{wang2024pvf} &
$94.38 \pm 0.22$ & $95.52 \pm 0.17$ & $94.56 \pm 0.42$ & $94.21 \pm 0.24$ & $98.82 \pm 0.21$ & $88.77 \pm 0.45$ \\

8  & Coat Lite Small \cite{wang2024pvf} &
$91.32 \pm 1.71$ & $93.14 \pm 1.33$ & $91.19 \pm 1.85$ & $91.48 \pm 1.66$ & $97.50 \pm 0.87$ & $82.66 \pm 3.41$ \\

9  & Swin Transformer V2 Tiny \cite{wang2024pvf} &
$92.40 \pm 0.78$ & $93.97 \pm 0.64$ & $92.40 \pm 0.74$ & $92.42 \pm 0.97$ & $98.25 \pm 0.26$ & $84.82 \pm 1.55$ \\

10 & Vision Transformer Small \cite{wang2024pvf} &
$87.79 \pm 1.47$ & $90.23 \pm 1.09$ & $88.19 \pm 1.93$ & $87.50 \pm 1.29$ & $95.47 \pm 0.63$ & $75.68 \pm 2.95$ \\

11 & GEPFNet \cite{GUO2026104014} &
$95.01 \pm 0.58$ & $96.05 \pm 0.42$ & $95.01 \pm 1.10$ & $95.06 \pm 0.47$ & $98.83 \pm 0.38$ & $90.06 \pm 1.15$ \\

\midrule
12 & \textbf{JEFFNet{} (Ours)} &
\boldmath$97.53 \pm 0.35$ &
\boldmath$96.41 \pm 0.51$ &
\boldmath$97.49 \pm 0.34$ &
\boldmath$97.57 \pm 0.58$ &
\boldmath$99.09 \pm 0.14$ &
\boldmath$90.95 \pm 1.32$ \\
\bottomrule
\end{tabular}%
}
\vspace{2mm}

{\footnotesize\raggedright
\textit{Note:} Bold values in the table indicate the best results. All results show the average and standard deviation from 10 runs.\par}
\end{table*}

\begin{table*}[!t]
\centering
\caption{Comparison of existing methods and the proposed model on the ISM dataset in the 12-Class task.}
\label{tab:ism-12-class}
\resizebox{\textwidth}{!}{%
\begin{tabular}{c l c c c c c c}
\toprule
\textbf{No.} & \textbf{Models/Study} & \textbf{F1} & \textbf{Acc} & \textbf{Re} & \textbf{Pre} & \textbf{AUC} & \textbf{MCC} \\
\midrule
1  & AlexNet                  & $63.76 \pm 1.23$ & $79.49 \pm 0.80$ & $62.85 \pm 1.20$ & $65.82 \pm 1.36$ & $94.51 \pm 0.26$ & $71.35 \pm 1.20$ \\
2  & EfficientNetB0           & $69.75 \pm 1.42$ & $83.09 \pm 0.55$ & $67.99 \pm 1.50$ & $72.99 \pm 1.54$ & $96.36 \pm 0.27$ & $76.29 \pm 0.80$ \\
3  & MobileNetV4 Small        & $70.24 \pm 0.72$ & $83.34 \pm 0.50$ & $68.72 \pm 0.64$ & $72.92 \pm 1.32$ & \boldmath$96.92 \pm 0.19$ & $76.67 \pm 0.71$ \\
4  & ConvNeXtV2 Base          & $65.20 \pm 1.27$ & $80.73 \pm 0.75$ & $62.72 \pm 1.26$ & $69.65 \pm 2.30$ & $95.47 \pm 0.27$ & $72.83 \pm 1.06$ \\
5  & FastViT SA               & $64.44 \pm 0.87$ & $80.08 \pm 0.36$ & $63.18 \pm 0.94$ & $66.90 \pm 1.46$ & $95.74 \pm 0.18$ & $72.15 \pm 0.56$ \\
6  & ResNet50                 & $56.75 \pm 1.78$ & $74.97 \pm 0.76$ & $54.67 \pm 1.56$ & $61.48 \pm 2.47$ & $93.58 \pm 0.37$ & $64.57 \pm 1.04$ \\
7  & EfficientNetV2           & $69.49 \pm 0.83$ & $83.41 \pm 0.14$ & $67.67 \pm 1.07$ & $72.94 \pm 0.89$ & $96.17 \pm 0.16$ & $76.75 \pm 0.18$ \\
8  & Vision Transformer Small & $55.71 \pm 1.17$ & $75.23 \pm 0.72$ & $53.60 \pm 1.33$ & $59.62 \pm 1.40$ & $93.01 \pm 0.47$ & $64.94 \pm 1.04$ \\
9  & Swin Transformer V2 Tiny & $66.77 \pm 0.95$ & $82.34 \pm 0.30$ & $64.98 \pm 0.93$ & $70.58 \pm 1.15$ & $96.05 \pm 0.21$ & $75.22 \pm 0.43$ \\
10 & Coat Lite Small          & --                & --                & --                & --                & --                & --                \\
11 & GEPFNet                  & $71.32 \pm 0.72$ & $83.69 \pm 0.35$ & $68.81 \pm 1.12$ & \boldmath$75.41 \pm 1.12$ & $96.39 \pm 0.35$ & $76.96 \pm 0.35$ \\

\midrule
12 & \textbf{JEFFNet (Ours)}
& \boldmath$72.60 \pm 1.96$
& \boldmath$83.88 \pm 0.73$
& \boldmath$71.47 \pm 2.18$
& $74.35 \pm 2.19$
& $96.47 \pm 0.46$
& \boldmath$77.33 \pm 1.06$ \\
\bottomrule
\end{tabular}
}
\vspace{0.5em}

\footnotesize\raggedright
\textit{Note:} The values in the table are the means and standard deviations of 10 runs. Bold values indicate the best results. Baseline results are taken from \cite{GUO2026104014}, while JEFFNet{} denotes our proposed method.\par
\end{table*}

\begin{table*}[!t]
\centering
\caption{Comparison of existing methods and the proposed model on the ISM dataset in the 2-Class task.}
\label{tab:ism-2-class}
\resizebox{\textwidth}{!}{%
\begin{tabular}{c l c c c c c c}
\toprule
\textbf{No.} & \textbf{Models/Study} & \textbf{F1} & \textbf{Acc} & \textbf{Re} & \textbf{Pre} & \textbf{AUC} & \textbf{MCC} \\
\midrule
1  & AlexNet                  & $92.52 \pm 0.55$ & $92.53 \pm 0.55$ & $92.53 \pm 0.55$ & $92.53 \pm 0.55$ & $96.91 \pm 0.25$ & $85.07 \pm 1.11$ \\
2  & EfficientNetB0           & $93.68 \pm 0.31$ & $93.69 \pm 0.31$ & $93.70 \pm 0.32$ & $93.71 \pm 0.32$ & $97.89 \pm 0.09$ & $87.41 \pm 0.64$ \\
3  & MobileNetV4 Small        & $93.87 \pm 0.51$ & $93.88 \pm 0.51$ & $93.90 \pm 0.51$ & $93.91 \pm 0.51$ & $97.81 \pm 0.11$ & $87.81 \pm 1.02$ \\
4  & ConvNeXtV2 Base          & $92.72 \pm 0.52$ & $92.73 \pm 0.52$ & $92.74 \pm 0.51$ & $92.74 \pm 0.51$ & $97.19 \pm 0.22$ & $85.48 \pm 1.02$ \\
5  & FastViT SA               & $87.51 \pm 2.52$ & $87.54 \pm 2.47$ & $87.60 \pm 2.44$ & $87.91 \pm 2.06$ & $93.89 \pm 2.08$ & $75.50 \pm 4.49$ \\
6  & ResNet50                 & $88.79 \pm 1.24$ & $88.79 \pm 1.24$ & $88.82 \pm 1.24$ & $88.87 \pm 1.22$ & $94.99 \pm 0.90$ & $77.69 \pm 2.46$ \\
7  & EfficientNetV2           & $93.49 \pm 0.29$ & $93.49 \pm 0.29$ & $93.50 \pm 0.30$ & $93.50 \pm 0.30$ & $97.81 \pm 0.13$ & $87.01 \pm 0.59$ \\
8  & Vision Transformer Small & $86.81 \pm 0.75$ & $86.81 \pm 0.75$ & $86.84 \pm 0.74$ & $86.92 \pm 0.72$ & $93.55 \pm 0.79$ & $73.76 \pm 1.46$ \\
9  & Swin Transformer V2 Tiny & $93.16 \pm 0.32$ & $93.17 \pm 0.32$ & $93.18 \pm 0.32$ & $93.18 \pm 0.32$ & $97.57 \pm 0.18$ & $86.36 \pm 0.65$ \\
10 & Coat Lite Small          & --                & --                & --                & --                & --                & --                \\
11 & GEPFNet                  & $93.93 \pm 0.24$ & $94.01 \pm 0.35$ & \boldmath$93.94 \pm 0.25$ & $93.94 \pm 0.24$ & $97.76 \pm 0.11$ & $87.92 \pm 0.50$ \\

\midrule
12 & \textbf{JEFFNet (Ours)}
& \boldmath$94.69 \pm 0.59$
& \boldmath$94.78 \pm 0.56$
& $93.16 \pm 1.08$
& \boldmath$96.28 \pm 0.60$
& \boldmath$98.30 \pm 0.31$
& \boldmath$89.61 \pm 1.10$ \\
\bottomrule
\end{tabular}
}
\vspace{1mm}

{\footnotesize\raggedright
\textit{Note:} The values in the table are the means and standard deviations of 10 runs. Bold values indicate the best results. Baseline results are taken from \cite{GUO2026104014}, while JEFFNet{} denotes our proposed method.\par}
\end{table*}

\subsection{Ablation Study}

To analyze the contribution of the main components of the proposed framework, we conducted an ablation study on the PVF-10 10-class classification task. The results are reported in Table~\ref{tab:ablation}. All variants were evaluated using the same fixed train/validation/test split. To better understand the contribution of each component, the ablation study is organized into three parts: (1) the effect of using a semantically rich out-of-domain dataset for JEPA representation learning, (2) the impact of different fine-tuning strategies for the pretrained JEPA encoder, and (3) the contribution of the proposed multibranch JEFFNet architecture.

\subsubsection{Rich Out-of-Domain Semantics for JEPA Representation Learning}

The first part of the ablation study investigates the effect of the dataset used for StoP-JEPA pretraining. Since StoP-JEPA employs a ViT as its encoder, its representation quality generally benefits from pretraining on large-scale and semantically rich datasets which is not fully exploited using existing solar datasets such as PVF-10 and ISM. To evaluate this, the same StoP-JEPA configuration was considered in two settings: one trained from scratch on the PVF-10 dataset and another pretrained on ImageNet-1k corresponding to the first and second row of Table~\ref{tab:ablation} respectively. Using a linear probing protocol, the StoP-JEPA encoder trained from scratch on PVF-10 achieves an accuracy of 68.95\% and an F1-score of 65.56\%. In contrast, the ImageNet-1k-pretrained StoP-JEPA encoder achieves 85.02\% accuracy and an F1-score of 82.70\%. This substantial improvement demonstrates that learning semantic representations from a large and semantically rich domain such as ImageNet-1k yields substantially better transferable representations than learning solely from the relatively small solar dataset such as PVF-10. Based on this observation, all subsequent experiments employ the ImageNet-1k-pretrained StoP-JEPA encoder.

\subsubsection{Fine-Tuning Strategies in a Single-Branch JEPA Scenario}

The second part of the ablation study examines how the pretrained StoP-JEPA encoder should be adapted to the downstream PV fault classification task. Three strategies are evaluated. In the \emph{Frozen} strategy corresponding to the second row of Table~\ref{tab:ablation}, the self-supervised pretrained encoder remains fixed throughout supervised training and only the classifier head is optimized. In the \emph{Unfrozen} strategy corresponding to the third row of Table~\ref{tab:ablation}, both the encoder and classifier head are optimized jointly from the beginning of supervised training allowing the representations to adapt to the new solar dataset domain. Finally, in the \emph{Staged} strategy corresponding to the fourth row of Table~\ref{tab:ablation}, only the classifier head is trained while the encoder remains frozen for the first few epochs. The encoder is then unfrozen, and the entire network is fine-tuned jointly for the remaining epochs. The results show that the Frozen strategy achieves 85.02\% accuracy and an F1-score of 82.70\%, while jointly training the entire network from the beginning increases the accuracy to 88.09\% but does not improve the F1-score, which remains at 82.41\%. In contrast, the proposed Staged strategy substantially improves all performance metrics, achieving 93.86\% accuracy and 93.69\% F1-score. These results indicate that gradually adapting the pretrained JEPA representations to the downstream task provides a more effective transfer strategy than either keeping the encoder permanently frozen or updating the weights of the entire network from the beginning. Therefore, the staged fine-tuning strategy is adopted in the proposed framework elaborated in the following subsection.

\subsubsection{JEFFNet: A Staged Multibranch Architecture using rich JEPA Representations}

The final part of the ablation study corresponding to last row of Table~\ref{tab:ablation} which investigates whether the semantic representations learned by StoP-JEPA in a self-supervised manner can be complemented with supervised convolutional features. To this end, we augment the staged JEPA architecture with an EfficientNetV2-S branch, producing the proposed multibranch JEFFNet architecture as shown in Fig~\ref{fig:jeffnet}. The resulting model achieves the best overall performance, reaching an accuracy of 94.77\%, an F1-score of 93.84\%, a precision of 95.86\%, an AUC of 99.70\%, and an MCC of 93.80\%. Although the single-branch staged JEPA model attains a slightly higher recall, JEFFNet performs better across the remaining evaluation metrics. These results suggest that the semantic representations learned by StoP-JEPA and the discriminative convolutional features extracted by EfficientNetV2-S provide complementary information, and that their combination yields the most reliable model for fine-grained PV fault classification.

\subsection{Experimental Comparisons}

The proposed framework was evaluated on two datasets: PVF-10 and ISM. For PVF-10, we report results in two settings: \textit{(i)} the original 10-class fine-grained PV fault classification task, and \textit{(ii)} a derived 2-class task in which the \textit{healthy} category is treated as one class and the remaining nine fault categories are grouped into a single \textit{faulty} class. Similarly, for the ISM dataset, we evaluate the models in the original 12-class classification setting and in a derived 2-class healthy-versus-faulty setting. In both datasets, the 2-class results were not obtained by training separate binary classifiers. Instead, they were derived directly from the multiclass model outputs by merging all fault categories into a single faulty class. This provides an additional evaluation of model behavior in a practically important PV inspection scenario, where the main objective is often to determine if a module is healthy or faulty. Tables~\ref{tab:pvf10-10-class} and~\ref{tab:pvf10-2-class} summarize the results on PVF-10. In the 10-class setting, JEFFNet{} achieves the best F1-score, recall, precision, and AUC among the compared methods. Compared with GEPFNet, the proposed model improves the F1-score by 0.74\%, recall by 0.78\%, precision by 0.94\%, and AUC by 0.26\%. GEPFNet obtains slightly higher accuracy and MCC, with margins of 0.31 and 0.36\%, respectively. These results indicate that although the accuracy difference is small, JEFFNet{} provides stronger performance in class-sensitive metrics, which is important for PVF-10 due to class imbalance and visually similar fault categories. In the derived 2-class PVF-10 setting, JEFFNet{} achieves the best performance across all reported metrics. Compared with GEPFNet, it improves the F1-score by 2.52\%, accuracy by 0.36\%, recall by 2.48\%, precision by 2.51\%, AUC by 0.26\%, and MCC by 0.89\%. The improvements in F1-score and recall are particularly important for practical PV inspection, since higher recall indicates that fewer faulty modules are missed during healthy-versus-faulty screening. The confusion matrix of the best-performing JEFFNet model on the PVF-10 test set is shown in Fig. \ref{fig:cm-best}. Overall, the matrix exhibits a strong diagonal structure, indicating that most defect categories are correctly identified. Four classes, namely \textit{break} and \textit{substring open circuit}, are classified without error, while \textit{shadow} and \textit{short circuit panel} also achieve recognition rates above 96\%. The largest source of confusion occurs for the \textit{debris cover} class, where a small number of samples are misclassified as \textit{healthy panel} (12.20\%) and \textit{junction box heat} (2.44\%). Minor confusion is also observed between \textit{healthy panel} and visually similar fault categories, particularly \textit{debris cover} (2.65\%) and \textit{bottom dirt} (1.99\%), while \textit{junction box heat} is occasionally predicted as \textit{healthy panel} (5.00\%). In addition, a small proportion of \textit{string short circuit} samples are classified as \textit{hot cell} (14.29\%), suggesting that these two fault types exhibit similar thermal signatures in some cases. Nevertheless, the limited number of off-diagonal entries demonstrates that JEFFNet effectively separates most fault categories despite the fine-grained nature of the PVF-10 dataset.

Tables~\ref{tab:ism-12-class} and~\ref{tab:ism-2-class} report the corresponding results on the ISM dataset. In the 12-class setting, JEFFNet{} achieves the best accuracy, F1-score, recall, and MCC. Compared with GEPFNet, the proposed model improves accuracy by 0.19\%, F1-score by 1.28\%, recall by 2.66\%, and MCC by 0.37\%. GEPFNet obtains higher precision by 1.06\%, while MobileNetV4 Small obtains the highest AUC, exceeding JEFFNet{} by 0.45\%. Overall, the results show that the proposed fusion strategy generalizes beyond PVF-10 and remains effective on the more challenging ISM dataset. In the derived 2-class ISM setting, JEFFNet{} achieves the best performance in most metrics. Compared with GEPFNet, it improves accuracy by 0.77\%, F1-score by 0.76\%, precision by 2.34\%, AUC by 0.54\%, and MCC by 1.69\%. GEPFNet obtains a higher recall by 0.78\%. Therefore, while GEPFNet is slightly more sensitive in detecting faulty samples, JEFFNet{} provides a stronger overall balance between fault detection and false-alarm reduction, as reflected by its higher F1-score, precision, AUC, and MCC. Fig. \ref{fig:ism-cm-best} presents the confusion matrix of the best-performing JEFFNet model on the ISM dataset. Compared with PVF-10, the ISM dataset exhibits substantially greater inter-class confusion due to the larger number of categories and the greater visual similarity among several anomaly types. Nevertheless, JEFFNet maintains high recognition rates for dominant classes such as \textit{Offline-Module} (100.00\%), \textit{Cell-Multi} (96.50\%), and \textit{Hot-Spot} (96.00\%). The most challenging categories are \textit{Cell} (45.00\%), \textit{Hot-Spot-Multi} (57.45\%), and \textit{Soiling} (58.14\%), which are frequently confused with semantically related classes. In particular, \textit{Cell} is often misclassified as \textit{Hot-Spot-Multi} (25.00\%) and \textit{Soiling} (20.00\%), while \textit{Diode} and \textit{Vegetation} are commonly predicted as \textit{Cell-Multi} (19.28\% and 16.04\%, respectively). Similarly, \textit{Cracking} is occasionally confused with \textit{Cell-Multi} (16.00\%), and \textit{No-Anomaly} exhibits moderate confusion with \textit{Diode-Multi} (13.41\%) and \textit{Soiling} (5.49\%). Despite these challenging cases, the overall diagonal dominance of the confusion matrix is consistent with the quantitative improvements reported in Table IV, demonstrating that JEFFNet provides robust multiclass discrimination while the remaining errors are primarily concentrated among thermally similar defect categories. In addition to classification performance, model size is an important factor for practical deployment. JEFFNet{} contains only 108.6M parameters, compared with 205.91M parameters for GEPFNet. This corresponds to a 47.2\% reduction in the number of parameters while maintaining competitive or superior performance across both datasets. These results suggest that the proposed fusion design improves the trade-off between predictive performance and model complexity, making it more suitable for scalable PV fault inspection systems.

\section{Conclusion}

In this work, we introduced JEFFNet, a multibranch architecture that combines JEPA-based self-supervised representation learning with EfficientNetV2-S-based supervised feature extraction for PV fault classification from thermal IR imagery. Our ablation study shows that a staged fine-tuning strategy substantially improves the transferability of pretrained StoP-JEPA representations, while fusing these semantic representations with EfficientNetV2-S convolutional features further improves performance over the single-branch staged JEPA model. On the PVF-10 and ISM datasets, JEFFNet achieves state-of-the-art or highly competitive performance under both the original multiclass and derived binary classification settings. Notably, these results are achieved using only 108.6M parameters, compared with 205.91M parameters for GEPFNet, corresponding to a 47.2\% reduction in model size, demonstrating a favorable trade-off between predictive performance and model complexity. Several future directions can further extend this work. First, incorporating multimodal data, such as combining infrared imagery with RGB or other sensing modalities, could further improve detection performance, as suggested by recent multimodal PV inspection studies \cite{pinho2025anomaly}. Second, investigating more advanced JEPA-based architectures, such as DSeq-JEPA \cite{he2025dseq}, may enable more discriminative representation learning and improve downstream classification performance. Third, exploring alternative fusion mechanisms beyond feature concatenation, such as attention-based or gated fusion, may further improve the integration of semantic and convolutional representations. Finally, future work could explore model compression, inference-time efficiency analysis, and architectural simplification to enable efficient deployment on resource-constrained inspection systems.


\bibliographystyle{IEEEtran}
\bibliography{references}

\end{document}